# Deep Learning-Enhanced Analysis for Delineating Anticoagulant Essay Efficacy Using Phase Microscopy

**Sudhanshu Shrivastava[1], Mohit Rathor[1†], Devyani Yenurkar [2†], Shivam K. Chaubey[1], Sudip Mukherjee [2]\* and Rakesh K Singh [1]\***

[1]Laboratory of Information Photonics, Department of Physics, Indian Institute of Technology, Banaras Hindu University, IIT (BHU), Varanasi-221005, Uttar Pradesh, India
[2]School of Biomedical Engineering, Indian Institute of Technology, Banaras Hindu University, IIT (BHU), Varanasi-221005, Uttar Pradesh, India

\* Authors to whom any correspondence should be addressed.
† These authors contribute equally to this work.

Email: krakeshsingh.phy@iitbhu.ac.in and sudip.bme@iitbhu.ac.in

**Abstract**

The coagulation of blood after it is drawn from the body poses a significant challenge for hematological analysis, potentially leading to inaccurate test results and altered cellular characteristics, compromising diagnostic reliability. This paper presents a deep learning-enhanced framework for delineating anticoagulant efficacy *ex vivo* using Digital Holographic Microscopy (DHM). We demonstrate a label-free, non-invasive approach for analyzing human blood samples, capable of accurate cell counting and morphological estimation. A DHM with an automated image processing and deep learning pipeline is built for morphological analysis of the blood cells under two different anti-coagulation agents, e.g. conventional EDTA and novel potassium ferric oxalate nanoparticles (KFeOx-NPs). This enables automated high-throughput screening of cells and estimation of blood coagulation rates when samples are treated with different anticoagulants. Results indicated that KFeOx-NPs prevented human blood coagulation without altering the cellular morphology of red blood cells (RBCs), whereas EDTA incubation caused notable changes within 6 hours of incubation. The system allows for quantitative analysis of coagulation dynamics by assessing parameters like cell clustering and morphology over time in these prepared samples, offering insights into the comparative efficacy and effects of anticoagulants outside the body.

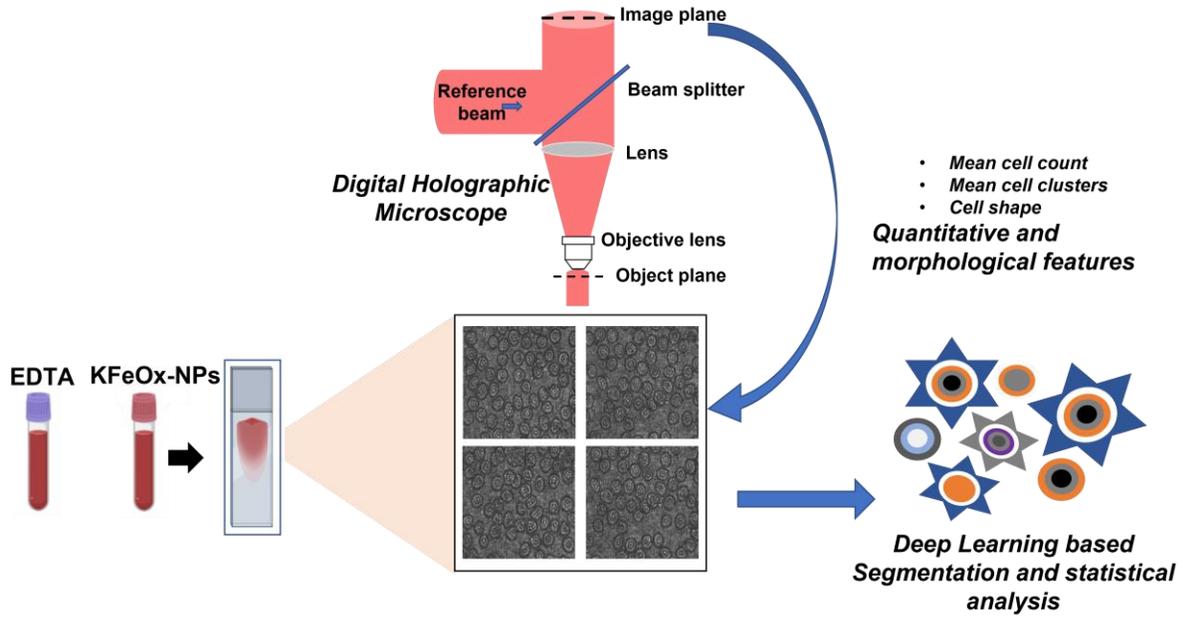

**Schematic 1:** Deep Learning-Enhanced Analysis for Delineating Anticoagulant Essay Efficacy Using Phase Microscopy

## 1. Introduction

Accurate hematological analysis is pivotal for diagnosing anomalies such as sickle cell anemia, thrombosis, diabetes, and malaria. Cell counting is a fundamental aspect of this analysis, which provides quantitative data on various blood cell populations and helps assess physiological conditions, disease progression, and treatment responses. Traditional cell counting and analysis methods have relied on electronic and optical techniques, each with inherent limitations[1]. One of the earliest automated approaches is the Coulter Counter, which measures changes in electrical impedance as cells suspended in an electrolyte pass through a micro-aperture [2-3]. This impedance-based method enables rapid counting but struggles with accuracy when cells cluster together. If multiple cells traverse the aperture simultaneously, the instrument registers a single, larger pulse (a "coincidence" event), causing it to miscount clustered cells [4]. Moreover, Coulter counters require cells to be in a conductive electrolyte solution to permit current flow. The reliance on an electrolyte medium introduces practical drawbacks: samples must be prepared in a specific ionic solution, which can risk contamination and demands careful system calibration. These constraints limit the Coulter counter's utility for routine blood cell analysis, especially in cell aggregates or debris samples. Subsequent advances turned to microscopy-based cell analysis, providing morphological information beyond mere counts. Optical microscopy techniques such as confocal[5] fluorescence microscopy and autofluorescence imaging [6] have been applied to blood cells to visualize cellular details and distinguish cell types. Confocal microscopy offers high-resolution optical sectioning, but it inherently captures only two-dimensional (2D) slices (typically up to a few hundred microns in depth), so multiple scans are needed to reconstruct 3D information [7]. Additionally, these imaging methods often depend on fluorescent dyes or labels to achieve contrast [8-9]. Staining cells with fluorescent markers can be toxic or perturbative to the cells, and the intense illumination required for imaging leads to photobleaching and phototoxic effects [10]. Even "label-free" modes like autofluorescence yield limited contrast and still fundamentally produce 2D projections. Thus, while optical microscopy provides richer information than impedance counts, it demands exogenous labels.

In recent years, DHM has emerged as a compelling alternative for label-free and phase imaging [11-15]. DHM is an interferometric quantitative imaging technique that records holograms of the sample [16-20] and computationally reconstructs the wavefront, yielding a quantitative phase contrast image related to the cell's thickness and refractive index distribution [21-23]. In effect, DHM can capture volumetric information without staining [24-27], i.e., a single hologram contains the data of different depths, allowing depth reconstruction [28-29] as well as live cell imaging [30-31]. This capability provides in situ 3D views of individual blood cells and their morphology, circumventing the need for mechanical z-scanning or fluorescent markers [25-27]. Furthermore, because it is label-free, DHM avoids toxic dyes and minimizes photodamage to cells. The high axial resolution of DHM makes it sensitive to sub-cellular morphological changes [12]. As a result, DHM-based cytometry can, in principle, quantify cell count and morphology in a single acquisition without the reagents or extensive sample preparation required by traditional optical methods.

The study of blood coagulation and the efficacy of anticoagulants are critical areas where advanced imaging can offer significant insights. Coagulation is vital, but its inappropriate initiation or inhibition in drawn blood samples can compromise diagnostic tests. Anticoagulants are routinely added to blood samples to prevent clotting and maintain sample integrity for analysis. However, different anticoagulants can have varied effects on cell morphology and functionality. The DHM can be used as a powerful, non-invasive means to observe these effects in detail. By providing quantitative phase data, DHM allows for the label-free monitoring of red blood cells (RBCs) morphology, aggregation (clotting), and density over time. In this paper, we built a DHM setup to test the performance of anticoagulants. The potassium ferric oxalate nanoparticle (KFeOx-NPs) is a new class of inorganic complex nanoparticle that was developed in our previous work, which prevents human blood coagulation and thrombosis in a mouse model [32].

This novel anticoagulant has been previously reported[38] to be non-toxic and biocompatible, with cytotoxicity and hemolysis assays showing cytocompatibility and hemocompatibility with HEK cell lines and human RBCs, respectively. Furthermore, the anticoagulant properties of the NPs were validated against commercial assays like prothrombin time (PT) and activated partial thromboplastin time (aPTT).

Here, the KFeOx-NPs were employed to compare their performance in blood with that of the commercial anticoagulant EDTA in terms of cellular morphology. Our approach involves acquiring holograms of treated blood smears at successive time points and analyzing changes in cellular characteristics to delineate anticoagulant efficacy and impact on cell structure.

However, adopting DHM for high-throughput blood cell analysis brings new data handling challenges. Manually annotating and analyzing such big imaging datasets is prohibitively time-consuming and labor-intensive [33]. This annotation bottleneck slows research and introduces variability, as different observers may produce inconsistent labels. Therefore, there is a critical need for automated analysis methods that can handle the scale and complexity of imaging-based blood cell data [34]. By leveraging these advanced segmentation networks, automatic blood cell segmentation in holographic or microscopic images is possible, enabling high-throughput cell counting and extraction of morphological features (e.g. cell size, shape, intracellular texture). The convergence of DHM with robust deep learning segmentation models offers a powerful route to overcome the limitations of traditional blood cell analysis methods, paving the way for more accurate, efficient, and scalable cytometry.

In this work, we use an automated data processing pipeline for analyzing blood cell coagulation using digital DHM upon administration of different anticoagulants [35-38]. This pipeline integrates Fourier fringe analysis [39-42] for phase image reconstruction and a U-net-based deep learning segmentation model (Cellpose 3.1 cyto3) [43]. By automating these steps, the system efficiently extracts morphological and optical features, such as cell density and coagulation rate, from large datasets generated by DHM. In this work, blood samples were prepared using two types of anticoagulants: the conventional EDTA, which prevents coagulation by chelating calcium ions [44-45] and KFeOx-NPs-based anticoagulant [38] maintains the structural integrity and optical properties of blood cells. While EDTA is effective at inhibiting clot formation, it can alter red blood cell morphology [46] and potentially distort quantitative phase measurements. In contrast, the nanoparticle-based anticoagulant is engineered to preserve cell morphology, offering an advantage for accurate diagnostic analysis. Over the course of the experiment, more than 400 holograms were acquired from blood smears at regular intervals using a high-resolution DHM system. The acquired holograms were processed through the pipeline to reconstruct phase images and accurately segment individual RBCs. The segmentation process, powered by the Cellpose 3.1 cyto3 model [43], enabled precise quantification of cell density and facilitated monitoring changes in cell distribution and clustering over time, which are key parameters for evaluating coagulation behavior.

## 2. Materials and Methods

### 2.1. Synthesis of Anticoagulant

The anticoagulant is an inorganic complex nanoparticle of potassium ferric oxalate (KFeOx-NPs), which was synthesized by a bottom-up method. Ferric sulphate (2.5 g) was added to barium oxalate (5 g) along with Poly(vinylpyrrolidone) (PVP) 250 mg in 60 ml of distilled water. The potassium oxalate (2.7 g) was added to this mixture after 30 min of stirring on a magnetic stirrer. After 15 min, 4 mL N/10 HCl was added and stirred for another 15 min. The filtrate of this mixture was then obtained after 3-4 hours of steam at 60-80°C. Further ethanol was added to the filtrate for centrifugation at 6500 RPM (25 °C) for 45 mins. The pellets formed were washed twice using ethanol and air dried. These dried particles (KFeOx-NPs) were characterized and treated with human blood samples.

### 2.2. Anticoagulation Assay

The anticoagulation property of KFeOx-NPs was tested using a clotting time assay. Blood containing anticoagulant EDTA, KFeOx-NPs, and untreated blood was used (100 μl). In a well plate, 10 μl of calcium chloride (0.2 M) was immediately added to the blood to accelerate the clotting process. These were done in triplicate and incubated at 37°C for 0, 2, 4, 6, 8, 10, and 12 hours. At respective time points, the absorbance of the washed solution was recorded at 540 nm using a SpectraMax M5 Microplate reader [47]. The Optical Density (OD) was significantly plotted using ANOVA in OriginLabs.

### 2.3. Sample preparation for microscopy

The human blood samples were used for all the experiments, as per the protocol approved by the Institute of Medical Sciences, Banaras Hindu University (Ref. No. ECR/526/Inst../UP/2014/RR-20, date- May 19, 2020). Blood samples of the same individual were collected in vials containing EDTA and KFeOx-NPs. The concentration of EDTA was chosen according to commercial standards (1.95 mg/ml). The KFeOx-NPs concentration was 2 mg/ml, which is close to the commercial EDTA concentration and was based on results from previous in vitro studies [38]. Smears were prepared on glass slides at each time point (0, 2, 4, 6, 8, 10, and 12 hours) using 5 µl of blood and allowed to dry for 7 mins. These smears were stained using the Leishman stain (2 min) and then washed and dried [48]. The prepared slides of both anticoagulants were further used for DHM imaging and analysis using high-tech pathology microscopy.

## 3. Technique

*3.1. Experimental setup for DHM and quantitative phase*

The schematic design of the experimental setup is shown in the figure 1. A vertically polarized light of wavelength 632.8 nm is emitted from a He-Ne laser (Newport, N-LHP-925). It enters the modified Mach-Zehnder interferometer through a spatial filter assembly and a half-wave plate (HWP) (Thorlabs, WPH10M-633) to make a diagonal (45°) polarized beam. The spatial filter consists of a microscope objective 10x, NA 0.25(Newport, M-10X), a pinhole of 25 microns, and a biconvex lens of focal length 200mm. The beam is focused by the microscopic objective lens and filtered by a pinhole aperture (A), which is placed at the back focal plane of the microscope objective, and then collimated by a biconvex lens placed at a focal length of 200 nm from the pinhole. HWP is placed at -22.5° with respect to the fast axis vertical. This collimated beam splits into two equal intensities through the beam splitter (BS1) and makes two arms of the interferometer. A beam transmitted from BS1, referred to as a reference, enters a precisely designed experimental scheme based on a triangular Sagnac geometry embedded with a telecentric geometry of L2 and L3. This triangular geometry comprised a polarizing beam splitter (PBS) and mirrors (M1, M2).

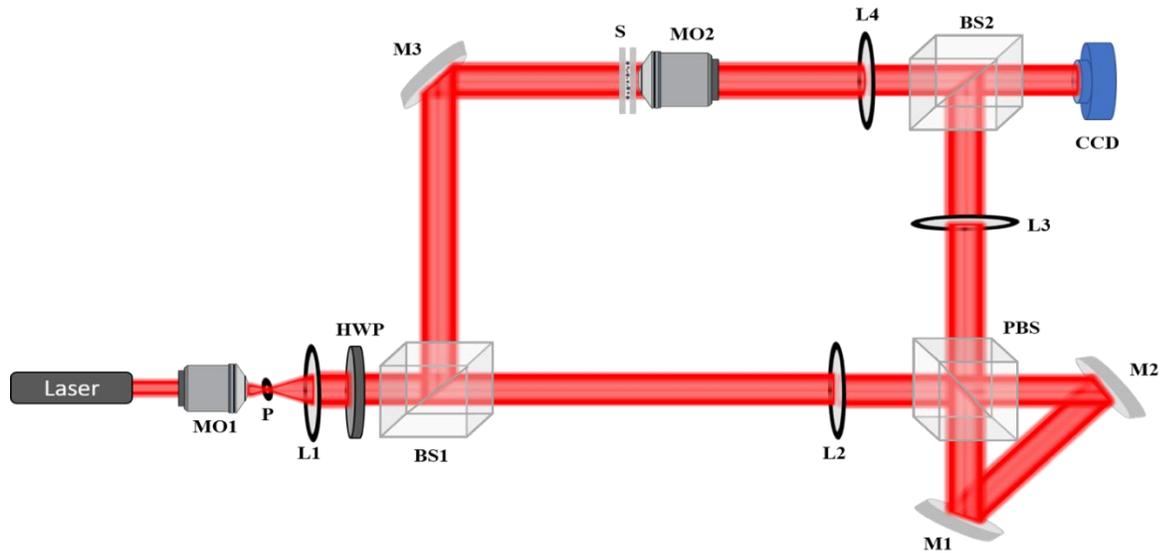

**Figure 1.** Schematic Experimental setup of a polarization-sensitive DHM system; HWP: half-wave plate; A: aperture; L1, L2, L3, and L4 are lenses; M1, M2, and M3 are mirrors; BS1 and BS2: beam splitter; PBS: polarizing beam splitter

The diagonally polarized reference beam splits into two orthogonal polarized beams after passing through PBS, which counter-propagate in a Sagnac geometry and exit from PBS as a pair of orthogonally polarized reference beams with tilts introduced by mirrors (M1 and M2). These tilts lead to spatial separation in the orthogonal polarization components at the back focal plane of lens L2 (which overlaps at the front focal plane of lens L3). Therefore, a reference beam is represented as,

$$E_R(r) = \hat{x}\, E_{Rx}(r) + \hat{y} E_{Ry}(r) \tag{1}$$

where $E_{Rx}(r) = exp(i\gamma_1 r)$ and $E_{Ry}(r) = exp(i\gamma_2 r)$. A beam reflected by the BS1 is folded towards the sample by a mirror M3. The beam from the sample plane scatters and is relayed at the detector plane by a 4*f* imaging system consisting of a 40x (NA 0.65) microscopic objective lens (MO2; Thorlabs, RMS40X).

Therefore, the object beam is represented as,

$$E_O(r) = \hat{x}\, E_{Ox}(r) + \hat{y}\, E_{Oy}(r) \tag{2}$$

Both reference and object beams propagate through the BS2, and a hologram is recorded at the charge-coupled detector (CCD) plane. The CCD is from Prosilica GT 1920 and has a pixel size of 4.54 µm. Therefore, the orthogonal polarized components of the sample are recorded, and the hologram is represented as,

$$I(r) = |E_o(r) + E_R(r)|^2 \tag{3}$$

The experimental design spatial resolution of the system is calculated as:
$$spatial\ resolution = \frac{0.61 \times \lambda}{NA}$$
$$= \frac{0.61 \times 0.6328}{0.65} \mu m = 0.6 \mu m$$

**Test and Validation the accuracy with Known Samples.** As shown in Figure 3, we used a polarizer and quarter wave plate (QWP) as standard and theoretically known samples to test the quality of the reconstruction of the orthogonal polarization components from our developed PDHM. Using orthogonal complex fields, we measured the polarization state at zero degree orientation angle of the polarizer with respect to pass axis horizontal and zero degree orientation angle of the QWP with respect to fast axis horizontal. Figures 2 (a),(b) and (c),(d) show the amplitude and phase information of the orthogonal polarization components for the polarizer oriented at 0° with respect to the pass axis horizontal. In this case, the theoretical output of the polarizer should be unit amplitude and zero phase in x-polarized Component, while y-polarized component should have zero amplitude with an arbitrary phase. The theoretical predictions show excellent quantitative agreement with the experimental measurements, thereby confirming the accuracy of the proposed approach. Figure 2(e)–(f),and (g)–(h) present the amplitude and phase distributions of the orthogonal polarization components for the quarter-wave plate (QWP) oriented at 0° with respect to the horizontal fast axis. Since the input beam is diagonally polarized, the theoretical expectation is a uniform unit amplitude for both polarization components. For the QWP oriented at 0°, the phase should be π/2 (1.57 rad) for the x-polarized component and zero for the y-polarized component. The experimental observations closely reproduce these theoretical values, demonstrating the reliability and precision of the system in polarization-resolved imaging.

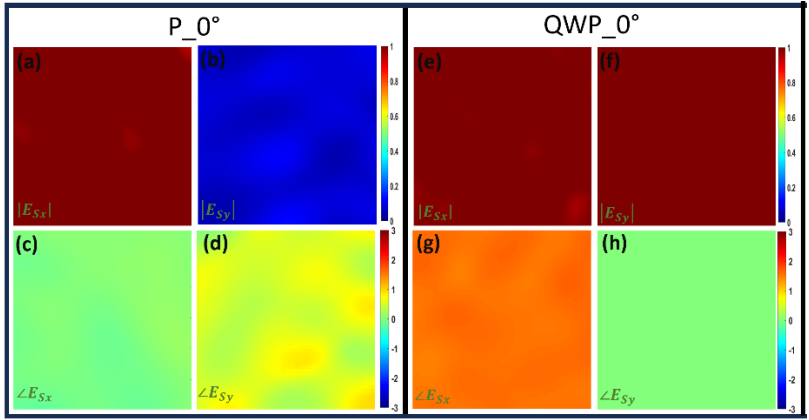

**Figure 2:** Amplitude and phase information of orthogonal polarization components for **(a)–(d)** a linear polarizer oriented at 0°, and **(e)–(h)** a Quarter-Wave Plate (QWP) oriented at 0°. The polarizer results **(a–d)** confirm the theoretical output of unit amplitude/zero phase for the $x$-component and zero amplitude for the $y$-component. The QWP results **(e–h)** show the expected uniform unit amplitude and a $\pi/2$ phase shift for the $x$-component relative to the $y$-component.

Fourier fringe analysis is used to reconstruct the quantitative phase images from the hologram. To recover the phase images of the cells from the hologram, we numerically reconstructed the hologram using our newly developed pipeline for the Fourier fringe analysis process. The novelty of this pipeline lies in its unique algorithm for determining the location and window size of the modulation peak in the Fourier spectrum, which enables truly automated peak finding without hardcoded parameters. This approach offers distinct advantages over other state-of-the-art methods: Unlike PyDHM [49], which uses a computationally expensive Cost Function Search [52], our method employs a simpler search for local maxima validated by symmetry analysis. Furthermore, our pipeline can handle any even number of modulation peaks, whereas PyDHM assumes only +1 and -1 peaks. Unlike PyHoloscope [50], which assumes a hardcoded radius for the AC peaks (e.g., 1/3rd of the modulation position), our method finds the peak extent through derivative analysis of the Fourier spectrum's intensity, avoiding fixed parameters. Unlike PyQPI [51], which uses a fixed limit parameter to determine peak width, our approach dynamically adapts to the data's characteristics. This prevents the under- or over-estimation of the peak width that can occur with a fixed limit when data varies between datasets. The Fourier Fringe analysis is done through automated processing of the hologram, and this numerical pipeline consists of several steps: Fourier transform, modulation peak identification and separation, peak centering, and inverse Fourier Transform. The Fourier spectrum of each hologram is first analyzed for global maxima. As shown in Figure 3(b), the standard maximum filter is applied first to the Fourier spectrum to detect the global maxima, and the brightest peaks found are compared through symmetry analysis. The modulation peaks are then confirmed by comparing the radius and the inclination of the points in the opposite quadrants of the spectrum. To automatically determine the extent of modulation peaks in the Fourier space, we have built a novel approach for finding the maxima along the 1-dimensional intensities of the spectrum along lines containing the modulation peaks. The region between the two consequent local maxima (the DC peak and the modulation peak) is selected, with the first maxima with the highest value in the derivative of the intensities after global minima in intensity, and the maximum of the two extents (along x and y axes) is selected Figure 3(c). This practice is repeated for all the modulation peaks. For this work, the algorithm yields four peaks for the hologram recorded for a diagonally polarized light as per carrier frequencies of the reference, with each pair of symmetrically opposite peaks about a particular polarization of light. The algorithm automatically selects the brightest peak and crops the region around it with the extent found earlier. This selected region is copied to an empty array of the dimension of the original spectrum, and the inverse Fourier transform of the resulting array yields the reconstructed complex field of the cells, in this case, the ROI of the RBC sample. Since the quantitative phase images of the RBC are the same for both orthogonal polarization components, i.e., non-birefringent response. Therefore, for further processing and evaluation of the coagulation in different environments, we selected phase images of only one polarization component for further analysis. Nevertheless, a similar analysis can also be performed for another polarization component if desired.

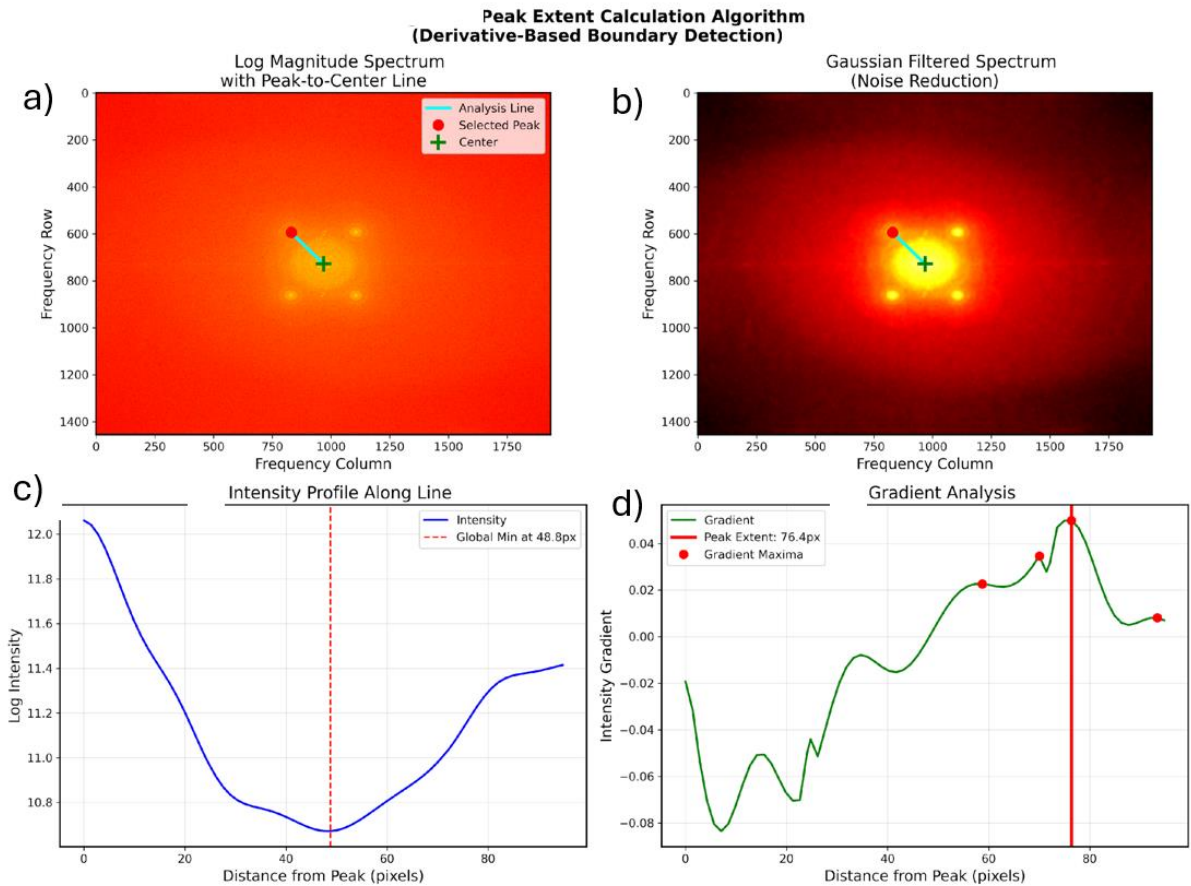

**Figure 3: Peak Extent Calculation Algorithm.** (**a**) Log magnitude spectrum with an analysis line (cyan), selected peak (green crosshairs), and its center (red dot). (**b**) Gaussian-filtered spectrum for noise reduction. (**c**) Intensity profile extracted along the analysis line, showing a global minimum. (**d**) Gradient analysis of the intensity profile, where the peak extent is identified by the maximum positive gradient after the global minimum, indicating the spectral boundary.

We have automated the process using a Python package, sequentially processing each hologram in our collected dataset to produce high-contrast phase images. The process takes around 137 minutes on a single HPC Param Shivay GPU node with a dataset of over 400 holograms. Figure 4 shows the reconstruction of complex information from recorded holograms of EDTA and KFeOx-NPs treated blood samples using Fourier fringe analysis. Thus, we use phase information to analyze anticoagulation properties further by the deep learning cell segmentation approach. These phase maps visualize optical path length variations across individual RBCs, revealing information about cell thickness, morphology in a label-free manner. Examining the reconstructed phase images from samples treated with EDTA and KFeOx-NPs uncovers distinct qualitative differences in cell appearance over time. In both cases, early time points (0–4 hours) display well-defined, circular RBCs with homogeneous phase contrast. As time progresses, EDTA-treated cells exhibit subtle distortions in their phase profiles. Instead of the typical biconcave disk shape, cells appear slightly elongated or asymmetrical, and local phase intensity variations emerge where cells contact one another. By later time points (8–12 hours), EDTA samples show dense regions of overlapping phase signals, indicating clusters of cells with poorly resolved boundaries. Conversely, KFeOx-NP–treated RBCs maintain circular outlines with uniform phase contrast throughout the entire 12-hour. Phase contours remain smooth and distinct, with minimal overlap even when cells lie close. These observations suggest that KFeOx-NPs preserve native cell morphology and inhibit aggregation more effectively than EDTA.

Spatially, the phase images highlight several features before segmentation: individual cell boundaries, variations in phase intensity that correspond to cell thickness (e.g., brighter centres indicating thicker midsections of biconcave cells), and regions where multiple cells lie adjacent or partially overlap. In EDTA samples at later times, overlapping cells produce merged phase regions with ambiguous contours, whereas in KFeOx-NP samples, cells remain spatially separated with clear, nonoverlapping phase footprints. After phase reconstruction, each image is processed to separate individual cell regions from contiguous phase patterns produced by overlapping or adjacent RBCs. A deep learning–based segmentation model identifies boundaries even when phase contrast is subtle or cells partially overlap. Segmentation converts each phase image into binary masks, which enumerate clusters and individual cells (Algorithm 1), providing input for the time-course metrics and follow-up comparisons.

## 3.2. Cell Segmentation

The 'cyto3' model from Cellpose 3.1 is used to segment each cell present in the $m_i$ reconstructed phase image [43]. The model outputs the mask matrix $M$. This generated mask $M$, with masks of each cell present, detected in the sample $m_i, i \in \{1, 2, ...N\}, N = |M|$ is the number of red blood cells detected. We selected this model for segmenting red blood cells from phase images because of its robust ability to handle the unique challenges posed by optical phase microscopy, since the model is robust to noise and distortions that may occur, and RBC morphology, since the model has been trained on a diverse dataset of cellular samples. The model excels at segmenting diverse shapes with an average precision of 0.771 at IoU (Intersection over Union) of 0.5 and a median F1 score of 0.861 over different test sets, including biconcave and irregular forms, and effectively separates overlapping or clustered cells using its vector flow-based approach. Its strong pre-trained versatility on biological imaging datasets ensures high baseline accuracy, while its adaptability allows fine-tuning for our specific data. Additionally, its efficiency in processing large datasets enables high-throughput analysis, providing precise segmentations essential for reliable quantitative and morphological measurements of RBCs. This was crucial for processing large datasets of phase images, enabling high-throughput analysis with no manual intervention. It includes capabilities for image restoration, such as denoising, deblurring, and up-sampling, which improve the quality of the input images and enhance segmentation accuracy.

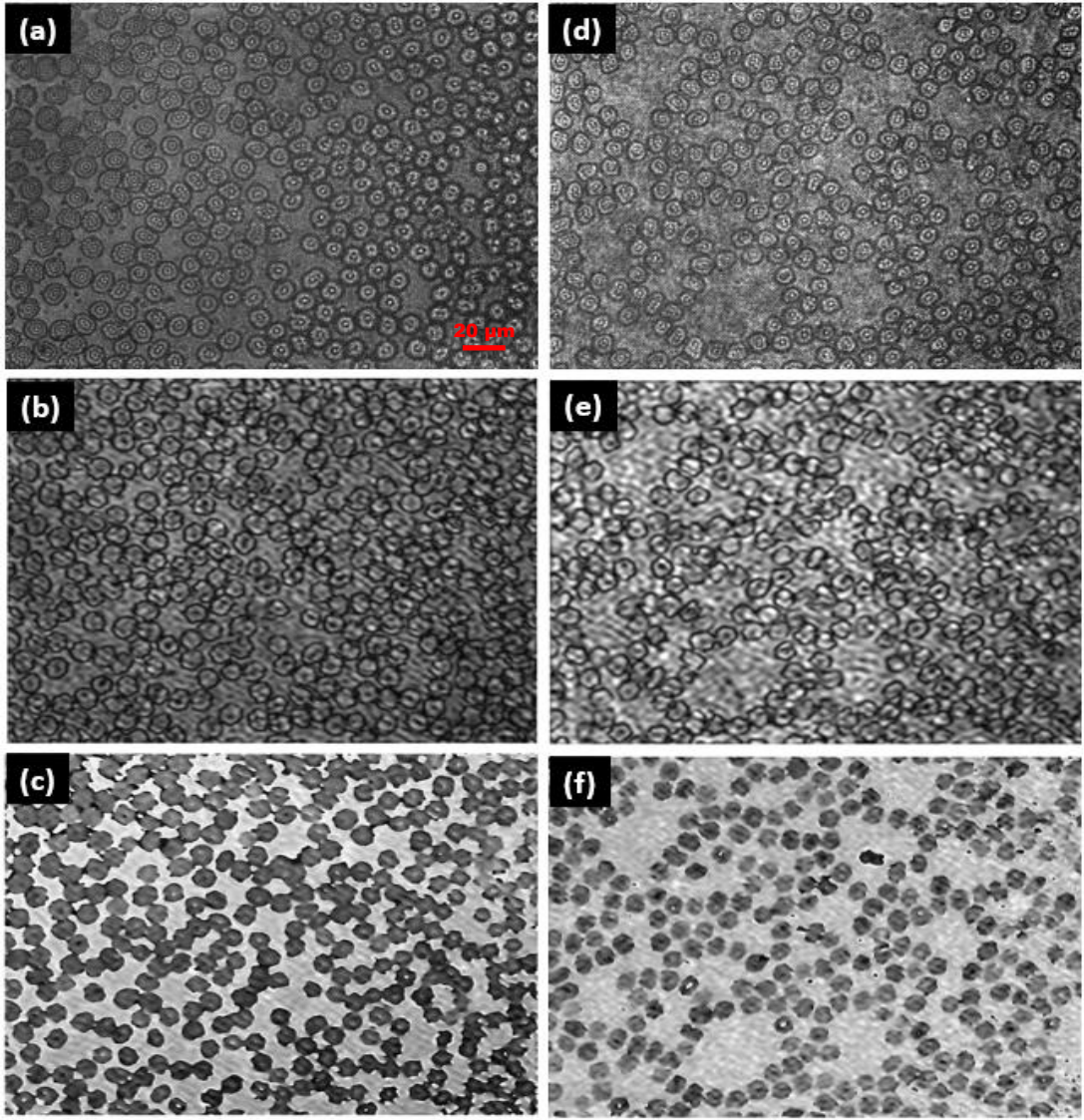

**Figure 4**. (a) and (c) show the recorded holograms, (b) and (e) show the reconstructed amplitude, and (c) and (f) show the reconstructed phase of blood samples treated with EDTA and KFeOx-NPs respectively.

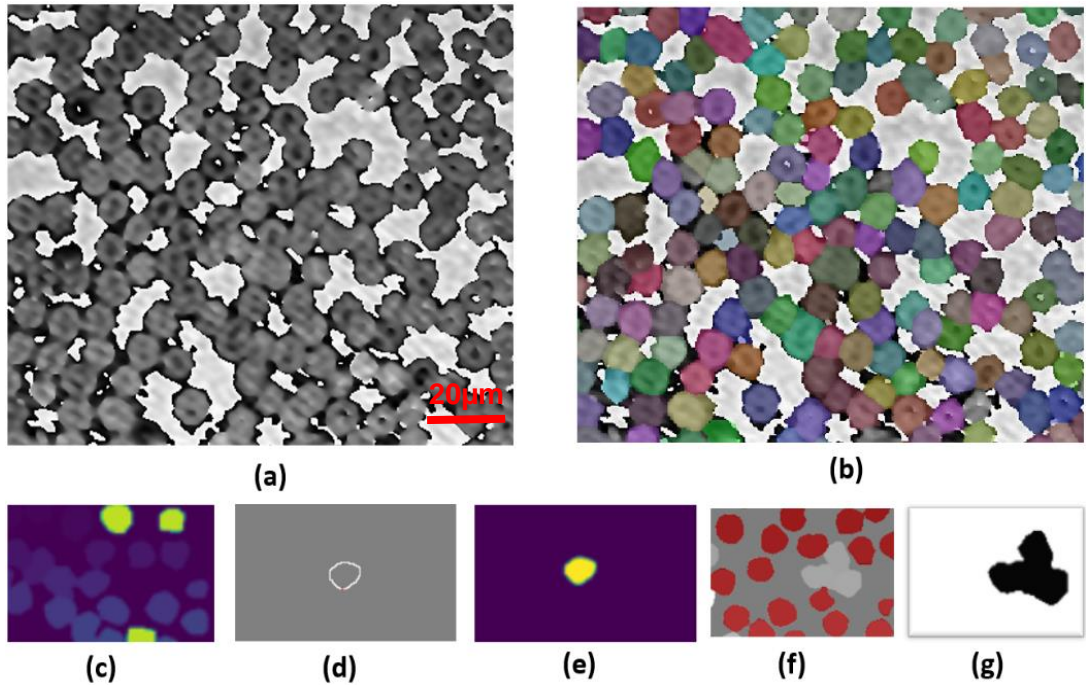

**Figure 5.** (a) Reconstructed optical phase image of red blood cells (RBCs) obtained through digital holographic microscopy (DHM); (scale bar: 20 µm). (b) Corresponding segmentation masks generated by the Cellpose 3.1 cyto3 deep-learning model, with each cell distinctly color-coded. (c-g) Step-by-step visualization of the automated cluster counting algorithm (Algorithm 1): (c) Initial segmentation output mask ; (d) Generation of vicinity mask by dilation around individual cell masks (Algorithm Steps 1 and 2); (e) Extraction of an individual RBC mask (Step 1); (f) Identification and marking of separated cells (red), which are subsequently removed from the mask (Steps 3 and 4); (g) Final binarized mask highlighting clustered RBC aggregates, used to compute the cluster count (Algorithm Steps 5–7).

### 3.3. Cluster count estimation

For each cell mask $m_i$, the corresponding dilation $s_i$ is calculated by convolving $m_i$ with the kernel matrix $k$ of dimension $(3,3)$. Then, the mask of pixels in the immediate vicinity of $m_i$ is calculated as $p_i = m_i - s_i$. The overlap factor $f$ is calculated by dividing the number of non-zero pixels on the application of $p_i$ to $M$, by $|p_i|$, i.e., the total number of pixels in the mask $p_i$. The factor $f$ ( we empirically determined it to be 0.15 for our dataset) is then compared by an empirically decided

---

**Algorithm 1** Algorithm for Cluster and Cell Separation

**Require:** Set of individual cell masks: $m_i$ (for each cell).
**Require:** Kernel matrix: $k$ of dimension $3 \times 3$.
**Require:** Full mask of cells: $M$.
1: **Step 1: Dilation of Each Cell Mask**
2: **for** each cell mask $m_i$ **do**
3:   Perform convolution with the kernel $k$ to calculate the dilation:
4:   $s_i = m_i \circledast k$ {where $\circledast$ denotes convolution}
5: **end for**
6: **Step 2: Calculate the Immediate Vicinity Mask**
7: **for** each $m_i$ **do**
8:   Calculate the mask of pixels in the immediate vicinity, denoted by $p_i$:
9:   $p_i = m_i - s_i$ {where $p_i$ represents the pixels in the immediate boundary around $m_i$}
10: **end for**
11: **Step 3: Calculate the Overlap Factor**
12: **for** each $p_i$ **do**
13:   Apply $p_i$ to the full mask $M$:
14:   $f = \frac{|p_i|}{\text{Number of non-zero pixels in } (p_i \cap M)}$ {where $|p_i|$ represents the total number of non-zero pixels in $p_i$}
15: **end for**
16: **Step 4: Compare Overlap Factor with Threshold**
17: **for** each overlap factor $f$ **do**
18:   **if** $f$ is below the threshold **then**
19:     Classify $m_i$ as a separate cell.
20:   **else**
21:     Classify $m_i$ as part of a cluster.
22:   **end if**
23: **end for**
24: **Step 5: Remove Separate Cells from the Full Mask**
25: **for** each separate cell $m_i$ **do**
26:   Remove $m_i$ from the full mask $M$:
27:   $M = M - m_i$
28: **end for**
29: **Step 6: Binarization**
30: Binarize the remaining mask $M$, such that all pixel values are either 0 or 1, representing absence or presence of clusters.
31: **Step 7: Count the Number of Clusters**
32: Calculate the number of connected components in the binarized mask $M$. The number of connected components corresponds to the number of cell clusters.
**Ensure:** The total number of clusters detected.

---

threshold value to determine whether $m_i$ is a separated cell or part of a cluster. All $m_i$ flagged as separate cells are then removed from the mask $M$. Finally, $M$ is binarized, and the number of connected components gives the cluster count. This algorithm is illustrated in figure 5.

## 4. Results

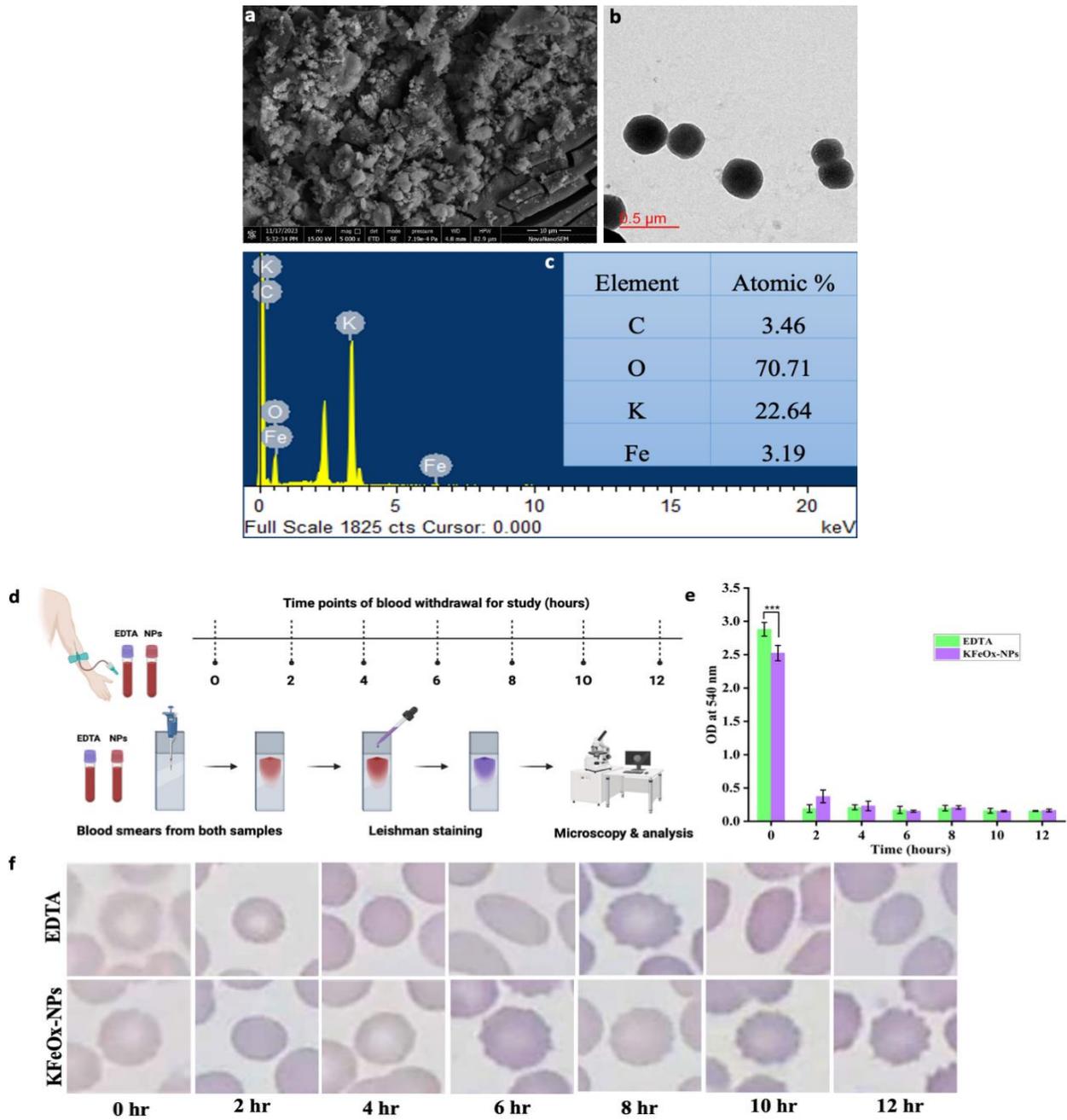

**Figure 6.** Characterization of KFeOx-NPs: (a) SEM image, (b) TEM imaging showing circular structure, and (c) elemental analysis of KFeOx-NPs, (d) Schematic representation of sample preparation, (e) Dynamic clotting time assay of KFeOx-NPs compared to EDTA, and (f) RBC images of blood containing EDTA and KFeOx-NPs at different time points.

An inorganic complex nanoparticle was synthesized to chelate the calcium in the blood, which alters the blood coagulation cascade. The structural and elemental analyses are shown in figure 6. Figure 6(a) displays scanning electron microscopy (SEM) images of KFeOx-NPs, and figure 6(b) confirms the circular structure of particles by transmission electron microscopy (TEM). Elemental analysis through Energy-dispersive X-ray analysis (EDS) in figure 6(c) shows the presence of iron, carbon, potassium, and oxygen. The anticoagulation potential of KFeOx-NPs was tested *in vitro* by dynamic clotting time using a $CaCl_2$-induced assay. Figure 6(e) illustrates blood clotting efficiency when incubated with anticoagulants, commercial EDTA, and KFeOx-NPs, and triggered with calcium chloride solution. Optical density (OD) difference at 2 hours suggests that the EDTA clots earlier than KFeOx-NPs. Next, we evaluated the cellular structure of RBCs from both the blood samples (containing EDTA and KFeOx-NPs) at 60X using the high-tech microscope from the PathLab (Sigtuple AI100). The microscopical analysis suggests a change in the morphology of the RBCs in blood containing EDTA after 6 hours. The circular structure of RBC was altered to an oval shape in the presence of EDTA, whereas in KFeOx-NPs containing blood, the morphology remains intact over a period of 12 hours, as shown in figure 6(f). This observation confirms the adverse effects of commercially used EDTA.

We quantitatively measured the phase structures of the RBC for different time gaps using the DHM. We compared the coagulation using the phase distributions and morphology. To study the coagulation property, we prepared the blood smear sample with EDTA and nano-particle-based anticoagulant at a successive gap of 0, 2, 4, 6, 8, 10, and 12 hours as illustrated in figure 6(d). The optical phase and a pseudo-3D map of the blood cell at different time gaps for different anticoagulant materials are shown in figure 7.

Figure 7 shows the quantitative phase map of the blood smear, which provides the blood's optical phase, depicting the blood cell's distribution inside the smear using our phase microscope. Note that the cell morphology changes with EDTA with time, and the cells are elongated and take elliptical shapes instead of keeping their circular structures. On the other hand, nanoparticles-induced blood cells are nearly circular, and they maintain their shape with time.

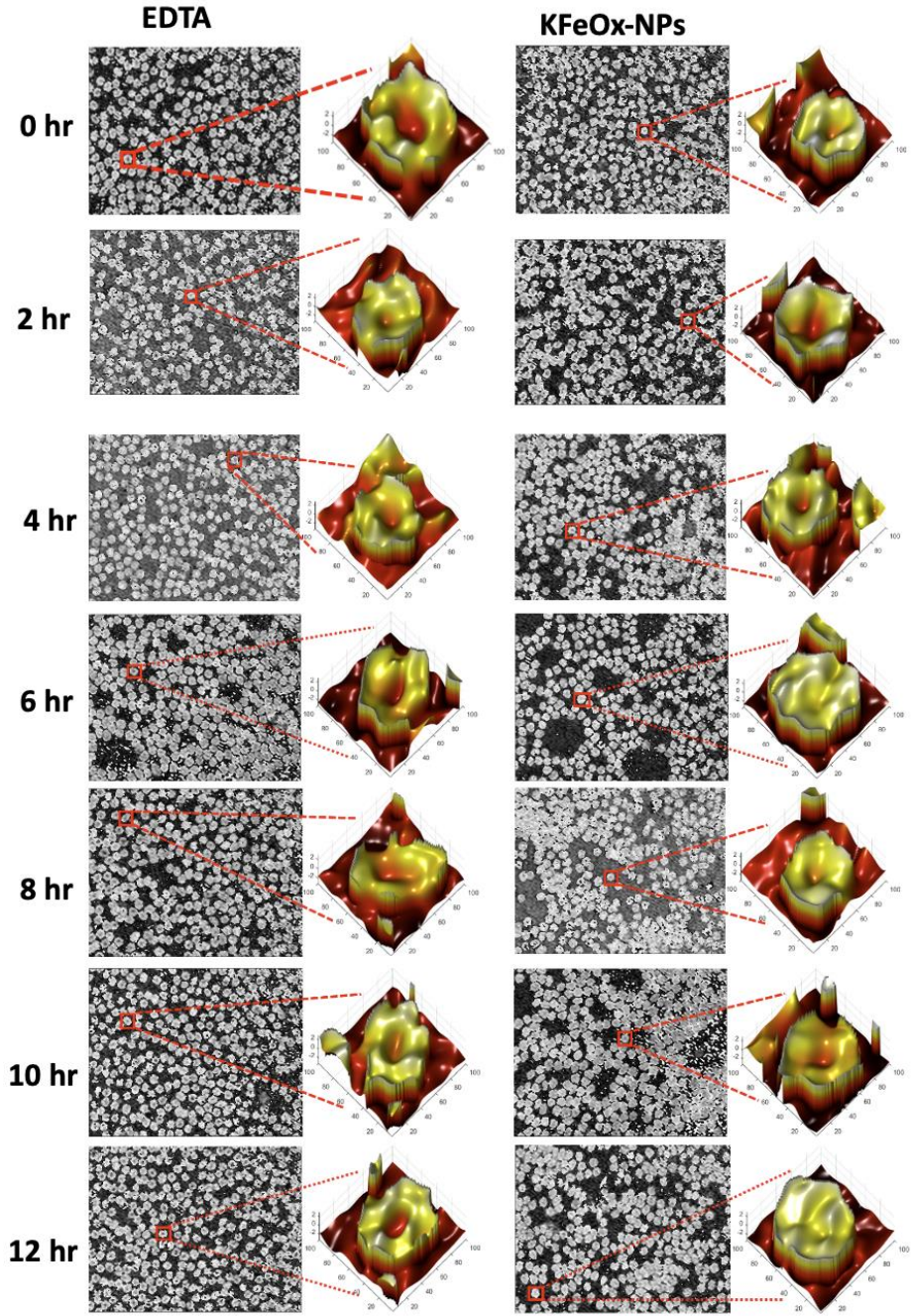

**Figure 7.** Experimental optical phase and pseudo 3D map result of a region of the whole slide image of blood samples subject to EDTA and KFeOx-NPs at successive times.

With the phase microscope, we quantitatively analyzed and presented the phase map of the RBC under different anticoagulants. Morphological changes in the RBC under different anticoagulants are given for the first time to the best of our knowledge, and we analyze how RBC evolves with time in the presence of EDTA and nanoparticle-based anticoagulants. Moreover, these experimental results demonstrate the efficacy of the nanoparticle-based anticoagulant compared to EDTA. Over time, analysis of the coagulation rates reveals differences between the two selected anticoagulants, namely EDTA and nanoparticles. The nanoparticle formulation exhibited superior anticoagulant properties most of the time, maintaining blood fluidity for extended periods compared to EDTA-treated samples.

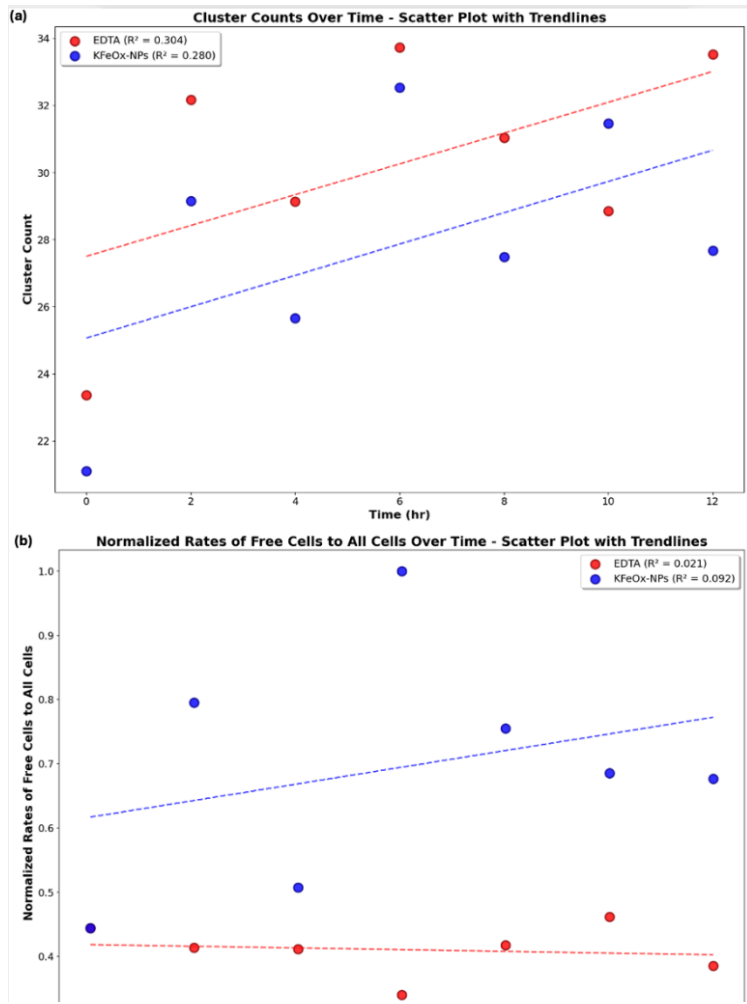

**Figure 8.** (a) A comparison between the coagulation rates of EDTA (Red) and KFeOx-NPs (Blue) infused whole blood samples with a linear fit. (b) A normalized trend chart depicting the rates of individual/free cells by total number of cells for EDTA (Red) and nanoparticles (Blue) infused whole blood samples

The time-course analysis of cell-cluster formation is shown in figure 8(a), which quantifies the number of multicellular aggregates ("clusters") observed per microscopic field as a function of incubation time. It illustrates the evolution of multicellular aggregation, with incubation time (hours) on the x-axis and the number of cell clusters per microscopic field on the y-axis. EDTA-treated samples exhibit a higher cluster count at most measured intervals than nanoparticle-treated blood, indicating more extensive cell–cell adhesion under the standard anticoagulant. In contrast, nanoparticle treatment consistently yields fewer aggregates throughout the 12-hour period.

Figure 8(b) presents the free-cell ratio, i.e., the proportion of single, unaggregated cells relative to the total cell population versus incubation time. The initial counts at t=0 hour have normalized the ratio to establish a fair point of comparison to counteract the differences in the initial concentrations of the cells in the different samples and regions of interest. EDTA treatment shows a relatively low free-cell proportion that gently declines over time, reflecting ongoing coagulation. Conversely, nanoparticle-treated samples maintain a markedly higher free-cell fraction, indicating more effective inhibition of aggregate formation. Taken together, these two complementary metrics consistently lower cluster counts and sustained high free-cell ratios under nanoparticle treatment demonstrate the superior anticoagulant efficacy of the nanoparticle formulation compared to EDTA. By suppressing the formation of multicellular aggregates and preserving single-cell suspensions over extended incubation, KFeOx-NPs represent a promising alternative for anticoagulant applications requiring prolonged blood fluidity.

## 5. Discussion

We used advanced deep learning methods combined with phase microscopy, especially DHM, to monitor the spatially resolved features of the RBC and identify the presence of anticoagulation in blood samples. While previous work has reported on the morphology of all cell types, including white blood cells (WBCs) and platelets, in this work, we focused the optical cell microscopy analysis specifically on RBCs. Ensuring sample integrity by preventing clot formation ex vivo is critical for dependable laboratory investigations and research. Anticoagulants are indispensable for preserving sample fluidity, yet their impact on blood cell integrity and the precise dynamics of coagulation inhibition require meticulous evaluation and observation. Conventional coagulation monitoring techniques, which often rely on time-dependent clotting tests and biochemical assays, can be intrusive, time-consuming, and may lack the real-time sensitivity and detailed morphological insights needed to assess anticoagulant effects at the cellular level fully. High-resolution viewing of blood samples is possible using DHM. This label-free imaging method monitors detailed morphological changes. It is important to note that we are not claiming to measure the refractive index (R.I.) directly; rather, we measure the quantitative optical phase, which is a multiple of the R.I. and the sample thickness. Deep learning-based analysis techniques alongside phase images data offer a novel way to enhance detection accuracy and sensitivity. This study also indicates significant differences in the morphology and behaviour of RBCs when treated with EDTA compared to the nanoparticle-based anticoagulant. The DHM system, coupled with the automated segmentation and statistical analysis pipeline, proved to be a robust tool for quantifying these differences. By providing label-free and non-invasive high-resolution imaging, the method allows for a detailed and quantitative analysis of blood sample coagulation dynamics. The study highlights the potential of integrating DHM with advanced machine learning models to improve the accuracy and efficiency of cell segmentation and provide deeper insights into the effects of different anticoagulants on blood cell integrity.

When applied to blood samples treated with either EDTA or KFeOx-NPs, the pipeline revealed marked differences in aggregation and morphology over a 12-hour period. EDTA-treated samples exhibited progressive RBC elongation and a plateau in free-cell ratios after approximately six hours, indicating both chelation-induced shape changes and persistent cluster formation. In contrast, KFeOx-NP–treated blood maintained circular morphology and consistently higher free-cell ratios, with cluster counts significantly lower than EDTA controls at all time points. These results suggest that KFeOx-NPs extend the fluidity window and minimize morphological artifacts that could confound downstream analyses.

Overall, this approach demonstrates that combining DHM with automated, AI-driven analysis can overcome the limitations of traditional methods. It allows for rapid, reproducible, and precise analysis of coagulation rates and cell density, making it a valuable tool for clinical diagnostics and biomedical research. The nanoparticle-based anticoagulant, in particular, shows promise in preserving the natural state of blood cells better than EDTA, thereby supporting more accurate assessments of blood cell behavior in various conditions. Beyond these specific anticoagulant comparisons, our pipeline's end-to-end processing speed and reproducibility significantly advance high-throughput cytometry. Processing over 400 holograms in under three hours on a single GPU node, without human intervention, highlights the annotation-efficient nature of our approach. Such scalability opens opportunities for large-scale studies of cell dynamics, ranging from morphological screening to real-time monitoring across diverse cell types and experimental conditions, all while preserving the native state of the specimen.

## 6. Conclusion

In this work, we demonstrated that integrating DHM with an automated, deep learning–driven analysis pipeline provides a powerful, label-free approach for high-throughput quantification of blood coagulation dynamics. By combining Fourier fringe analysis for robust phase reconstruction with a U-net–based segmentation model (Cellpose 3.1 cyto3), our system accurately extracted both morphological and optical features, such as individual cell counts, cluster formation rates, and free-cell ratios, from over 400 holograms in an end-to-end fashion. This automated workflow eliminates manual annotation bottlenecks and achieves reproducible measurements with minimal user intervention. This technique was used to examine the performance of conventional anticoagulants i.e. EDTA and a new class of anticoagulants i.e. KFeOx-NPs by analyzing the phase images of the RBCs. These results revealed that the KFeOx-NPs consistently maintained higher free-cell fractions and lower cluster counts over a 12-hour incubation period, indicating superior red blood cell aggregation inhibition. Moreover, phase-contrast imaging highlighted that KFeOx-NPs preserved native cell morphology more effectively than EDTA, underscoring their potential as a next-generation anticoagulant for diagnostic and research applications. By leveraging readily available computational resources and open-source software, this approach can be deployed for real-time monitoring of coagulation status and early detection of erythrocytic pathologies. Looking forward, the modular design of our DHM + AI framework lends itself to translation into point-of-care devices while broadening clinical utility.

## Acknowledgments

S.S. would like to acknowledge the financial support from the Department of Biotechnology (DBT)-BT/PR35557/MED/32/707/2019. and I-DAPT IIT(BHU) (Grant: R&D/SA/I-DAPT IIT(BHU)/PHY/23-24/04/495). We also thank Professor Dr. Arnab Sarkar in getting ethical approval for the use of human blood samples.

## Competing interests

The authors declare no competing financial interests.

## Ethical statement

All experiment were performed according to the protocol was approved by the ethical committee of the Institute of Medical Sciences, Banaras Hindu University (ref ECR/526/Inst./UP/2014/RR-20 date- May 19, 2020)